# Magnetically Induced Field Effect in Carbon Nanotube Devices

Georgy Fedorov [1], Alexander Tselev [2], David Jiménez [3], Sylvain Latil [4], Nikolay G. Kalugin [5], Paola Barbara [2], Dmitry Smirnov [1] and Stephane Roche [6]

[1] *National High Magnetic Field Laboratory, Tallahassee, FL 32310, USA*

[2] *Department of Physics, Georgetown University, Washington, DC 20057, USA*

[3] *Departament d'Enginyeria Electrònica, Escola Tècnica Superior d'Enginyeria, ETSE, Universitat Autònoma de Barcelona, UAB, 08193-Bellaterra, Barcelona, Spain*

[4] *Department of Physics, Facultes Universitaires Notre-Dame de la Paix, 61 Rue de Bruxelles, B 5000 Namur, Belgium*

[5] *Department of Chemistry, New Mexico Tech, Socorro, New Mexico 87801, USA*

[6] *Commissariat à l'Energie Atomique, DSM/DRFMC/SPSMS/GT, 17 rue des Martyrs, 38054 Grenoble, France*

## Abstract

Three-terminal devices with conduction channels formed by quasi-metallic carbon nanotubes (CNT) are shown to operate as nanotube-based field-effect transistors under strong magnetic fields. The off-state conductance of the devices varies exponentially with the magnetic flux intensity. We extract the quasi-metallic CNT chirality as well as the characteristics of the Schottky barriers formed at the metal-nanotube contacts from temperature-dependent magnetoconductance measurements.

KEYWORDS: carbon nanotubes; electronic transport; magnetoresistance; Aharonov-Bohm effect

The exceptional low-dimensionality and symmetry of carbon nanotubes (CNT) are at the origin of their spectacular physical properties governed by quantum effects. The nature of the CNT electronic spectrum is strongly dependent on its chirality $(n,m)$.[1] The numbers $(n,m)$ relate the wrapping vector that connects two matching sites on the nanotube's surface to the unit vectors of graphene lattice. A CNT with $n-m \neq 3i$, where $i$ is an integer, is semiconducting with an energy gap $\varepsilon_g \sim 1/r$ ($r$ is the radius of the CNT). Armchair nanotubes with $n=m$ are truly metallic,[2,3] while those with $n-m=3i$, $n \neq m$, are said to be quasi-metallic, owing to the presence of a small energy gap $\varepsilon_g \sim 1/r^2$ arising from curvature effects.[4]

Ajiki and Ando [5,6] predicted that an axial magnetic field would tune the band structure of a CNT between a metal and a semiconductor one, owing to the modulation of the Aharonov-Bohm (AB) phase [7] of the electronic wavefunctions. When a magnetic flux $\phi$ threads the nanotube cross section, the electron wavefunctions accumulate an additional magnetic field dependent phase factor (AB-phase), which results in a $\phi_0$-periodic modulation of the band gap ($\phi_0 = h/e$ is the flux quantum).[8-11] This phenomenon occurs due to peculiar topology of the graphene Fermi surface, and it is unique for carbon nanotubes. In Ref. 6, it is shown that, neglecting curvature induced effects, the field-dependent gap $\varepsilon_g(B)$ can be described for $\phi \leq \phi_0/2$ by a linear function:

$$\varepsilon_g(B) = \lambda |B - B_0|, \qquad (1)$$

where $B_0 = 0$ for all CNTs with $n-m = 3i$ and $B_0 = \phi_0/12\pi r^2$ otherwise. The coefficient $\lambda$ depends only on the CNT radius: [6]

$$\lambda \equiv d\varepsilon_g / dB = 3\pi a_{C-C} \gamma_0 r / \phi_0, \qquad (2)$$

where $a_{C-C}$ is the C-C bond distance, $\gamma_0$ is the nearest-neighbor interaction parameter. The validity of Eq. 2 has been confirmed experimentally by spectroscopic [12,13] and electric transport studies,[10-11,14-16] while signatures of $\phi_0$ periodic modulation of the band gap have been observed in large-diameter multiwall CNTs.[10,11,17]

In this Letter we report on magnetotransport measurements of devices made in the configuration of a standard CNT field-effect transistor (CNFET) [18] with a metallic or quasi-metallic single-walled CNT (SWNT). An exponential decrease of the device off-state conductance is observed under high axial magnetic fields up to room temperature. It is consistently described by the opening of an energy gap in the CNT electronic spectrum and modulation of the Schottky barriers at the CNT/metal interfaces. We also show that magnetotransport measurements provide a hallmark of the quasi-metallic nanotube chirality.

To fabricate devices shown in Figure 1a, single-walled nanotubes were grown on highly conductive n-doped Si substrates, capped with 400 nm of thermally grown $SiO_2$ and used as a back gate. CNTs were grown by a chemical vapor deposition method [19] from a $CH_4/H_2$ mixture using a Fe/Mo catalyst yielding a low density of nanotubes on the substrate. The catalyst was supported by mesoporous alumina islands of about 4 μm by 4 μm size. Each nanotube was contacted with two 50 nm-thick Pd electrodes patterned using

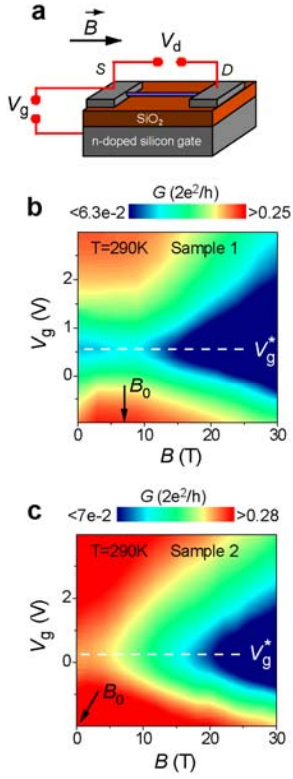

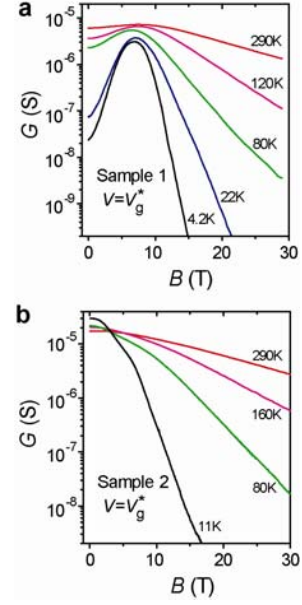

**Figure 1.** Tuning CNT devices characteristics by magnetic field: (a) schematic of a CNFET-type device, (b), (c) linear conductance $G = I/V_d$ of sample 1 and sample 2 at room temperature as a function of the gate voltage, $V_g$, and magnetic field $B$. The $G(V_g)$ characteristics were recorded at a small DC bias voltage $V_d \sim 1$ mV. The white dashed lines indicate $V_g^*$ values corresponding to the conductance minimum. Dark arrows indicate the values of $B_0$, where $\varepsilon_g(B)$ has a minimum.

conventional photolithography and deposited by RF-sputtering.

Measurements are performed on two devices, sample 1 and sample 2, made with individual single-walled nanotubes (SWNT) with diameters of 1.5 ± 0.5 nm and 2.6 ± 0.5 nm, respectively. Figure 1b,c displays 2D-plots of the device conductance, $G$, versus the gate voltage, $V_g$, and the magnetic field, $B$, applied parallel to the tube axis. At $B = 0$, sample 1 shows suppressed conductance in a certain gate voltage range, which can be interpreted by the presence of a small ~10 meV band gap in a quasi-metallic CNT. [4,20,21] The evolution of the transfer characteristic $G(V_g)$ suggests that the initially small band gap is progressively reduced as the magnetic field increases, to reach a minimum value at $B_0 \approx 6$ T. Above $B_0$, the band gap enlarges continuously with increasing $B$. In contrast, at zero magnetic field, the conductance of sample 2 remains almost constant versus the gate voltage, as expected for a truly metallic nanotube. When an axial magnetic field is applied, a region of suppressed conductance develops, indicating a monotonous increase of a band gap. Remarkably, under high axial magnetic fields both devices operate as CNFETs with the on/off conductance ratio exceeding $10^4$ at, e.g., 10 K and 20 T. Perpendicular magnetic fields do not significantly change the conductance in the studied gate voltage -10 V $\leq V_g \leq$ 10 V and temperature (4.2 K $\leq T \leq$ 290 K) ranges.

**Figure 2.** Off-state magnetoconductance of the CNT-based devices: Magnetoconductance curves $G(B)$ of sample 1 (a) and sample 2 (b) measured at $V_g = V_g^*$ at different temperatures. At $B > 10$ T the $G(B)$ curves of both samples appear as straight lines in the log-vs.-linear scale.

The axial magnetic field has the strongest effect on $G(V_g)$ at $V_g \sim V_g^*$ (off-state) corresponding to the conductance minima (Figure 1b,c). For both samples the value of $V_g^*$ changes with temperature, but remains magnetic field independent for $B > 10$ T. Figure 2 shows magnetoconductance curves $G(B)$ recorded at several temperatures between 4 K and 290 K and at $V_g = V_g^*$. At high enough fields, $B > 10$ T, the conductance of both samples decreases exponentially at all temperatures. This suggests thermally-activated carrier transport, $G \propto \exp(-\Delta(B)/k_B T)$ ($k_B$ is the Boltzmann constant), with an effective activation energy $\Delta(B)$ scaling linearly with the magnetic field. Qualitatively, this observation agrees with the predicted linear $\varepsilon_g(B)$ dependence, and therefore strongly favors the interpretation of our data in terms of the AB effect. Accordingly, the linear slope of the $\Delta(B)$ dependence, $\alpha = d\Delta(B)/dB$, is related to the rate of the energy gap growth in the magnetic field $\lambda = d\varepsilon_g/dB$. As shown below, the value of $\alpha$ depends on both the intrinsic CNT and device properties. The temperature dependence of $\alpha(T)$ provides information about the energy band profile of the CNT forming the conduction channel of our devices.

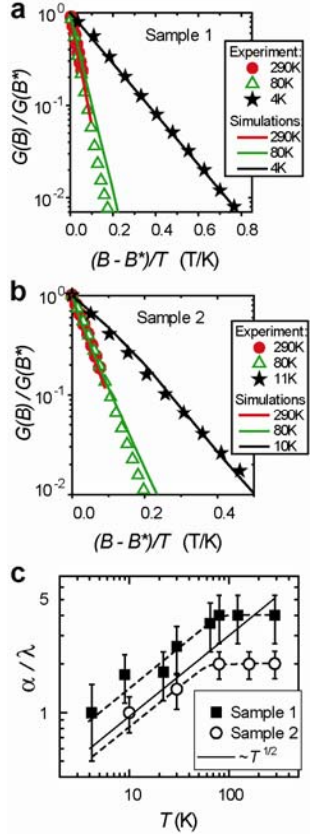

**Figure 3.** Exponential magnetoconductance of the CNT-based devices: (a, b) normalized conductance $G' = G(B)/G(B^*)$ of samples 1 and 2 as a function of the rescaled magnetic field $B' = (B - B^*)/T$, where $B^* = 12$ T and $B^* = 10$ T for samples 1 and 2, respectively. Solid lines show simulation results for the (19,10) CNT (a) and (19,19) CNT (b). For both samples, the $G'(B')$ curves obtained at $T > 80$ K collapse onto the same line indicating temperature-independent $\alpha$ values. (c) The ratio $\alpha(T)/\lambda$ as a function of temperature. The large errors bars are mostly due to systematic uncertainties in the values of $\lambda$ arising from errors in the CNT diameter measurements. The solid line highlights the $T^{1/2}$ power-law. Dashed lines are guides to the eye.

In Figure 3a,b the normalized conductance $G' = G(B)/G(B^*)$ of samples 1 and 2 are plotted as a function of the rescaled magnetic field $B' = (B - B^*)/T$, where $B^* = 12$ T and $B^* = 10$ T for samples 1 and 2 respectively. Figure 3c displays a log-log plot of $\alpha(T)$ extracted from the slope of these curves, normalized to values of $\lambda$ calculated using eq. 2. We find that the behavior of $\alpha(T)$ below approximately 80 K can be represented by a simple relationship $\alpha(T)/\lambda \propto T^{1/2}$, for both samples, and notably, it appears to be temperature-independent above that temperature. We note, that if the conductance of a nanotube is described by $G \propto \exp(-\varepsilon_g/2k_BT)$, as for an intrinsic semiconductor, one will get $\alpha = \lambda/2$ for all temperatures.

We begin our analysis by calculating the $\varepsilon_g(B)$ dependence for quasi-metallic and metallic SWNTs of different chiralities taking into account the curvature effects.

The electronic structure of CNTs is computed using the Slater-Koster tight-binding model with parameters from ref. 22. Within this approximation $\varepsilon_g(B)$ is found to be well described by equations (1) and (2). The hallmark of the curvature effects is a non-zero value of $B_0$ for $n \neq m$ that appears to be very sensitive to the $n/m$ ratio. For example, quasi-metallic CNTs with a diameter in the range from 1.4 nm to 1.6 nm have been found to have $B_0$ ranging from 5.7 T for a (13,10) CNT to 35.9 T for a (18,0) CNT. The remarkably strong dependence of $B_0$ on the chirality can be used to identify the helical symmetry of quasi-metallic CNTs. Given the uncertainty of our nanotube diameter measurements, one first assigns 53 different possible chiralities for sample 1 and 92 for sample 2. However, the theoretical calculations of $B_0$ allow us to reduce the choice of the chirality for sample 1 either to (13,10) with $B_0 = 5.7$ T or to (19,10) with $B_0 = 6.24$ T. For sample 2, with $B_0 = 0$ T, seven different truly metallic ($n = m$) CNTs with $16 \leq n \leq 22$ need to be considered. Neither diameter measurements nor magnetotransport data allow choosing any particular one of these seven chiralities. We model sample 2 by the (19,19) CNT with diameter of 2.58 nm, and further find that its calculated transport properties well reproduce experimental data.

To describe the electric transport in our devices under magnetic fields, we consider a CNT energy band diagram with $B$-dependent Schottky barriers that form at the nanotube/metal interface. The height of the Schottky barrier for electrons equals $\varphi_0 + \varepsilon_g(B)/2$, where $\varphi_0$ is approximated as the difference in the work-functions of Pd and CNT[23] ($\varphi_0 \approx 400$ meV). We assume that the band profiles, $E_{C,V}(y)$, have two characteristic length scales along the CNT direction $y$, as sketched in Figure 4a. The length scale of the band bending far away from the electrodes, $y_0$, is determined by the electrostatic influence of the gate[14] and is of the order of the SiO$_2$-layer thickness[24,25] $t_{ox} = 400$ nm. Close to the electrodes, the midgap energy is shifted from $\varphi_0$ down to $\varphi_1$ on the short length scale $y_1 \ll y_0$, thus inducing the formation of a thin tunneling barrier for electrons. The existence of such an additional short-scale interface barrier can be attributed to doping-induced effects, as already evidenced experimentally in ref. 26. Both parameters $\varphi_1$ and $y_1$ depend on the chemical environment and/or the particular geometry of the electrodes.

For a further semi-quantitative analysis of the device's magnetoconductance, charge transport through the nanotube is simulated using Landauer-Büttiker formalism, assuming ballistic conduction along the tube axis, whereas the energy band profiles $E_{C,V}(y)$ are described by:

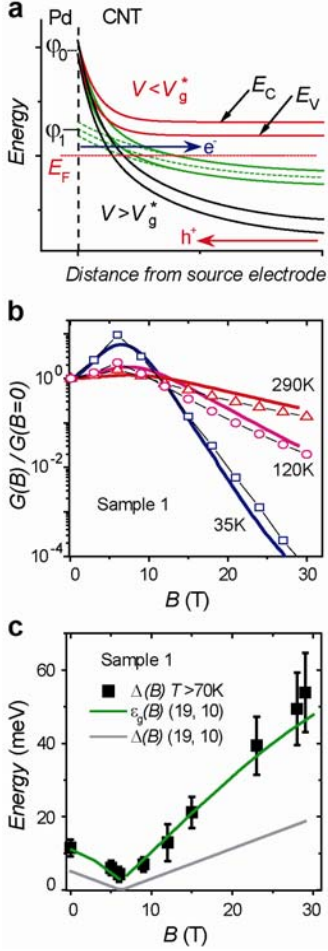

**Figure 4.** Magnetoconductance of CNT-based devices: intrinsic versus contact phenomena (a) Band profiles at different gate voltages. At negative gate voltages, there are no barriers for the carriers at the Fermi level. At large enough positive gate voltages, the electron transport through the nanotube occurs due to thermally-assisted band-to-band tunneling (blue arrow) and thermally-activated hole transport (red arrow) over the barrier in the bulk of the CNT. The gate voltage alters the height of the barrier for holes and the thickness of the tunneling barrier for electrons. (b) Conductance of sample 1 (solid lines), normalized to its value at zero magnetic field compared to simulations for the (19,10) CNT (symbols) (c) $\Delta(B)$ determined from the conductance Arrhenius plots for sample 1 (symbols) and (19,10) CNT (green line). Grey line: calculated dependence of the band gap $\varepsilon_g(B)$ of a (19,10) CNT on magnetic field.

$$E_{C,V}(y) = \varphi_1 \pm \frac{\varepsilon_g(B)}{2} + (\varphi_0 - \varphi_1)\exp\left(\frac{-y}{y_1}\right) - (\varphi_1 + e\chi(V_g, B))\left(1 - \exp\left(\frac{-y}{y_0}\right)\right) \quad (3)$$

where the midgap energy far from the electrodes $e\chi(V_g, B)$ is controlled by the gate voltage [27] with $\varepsilon_g(B)$ dependence obtained by the band structure calculations. To simulate the magnetoconductance curves $G(B)$, the minimum value of conductance as a function of $V_g$ is found at each magnetic field.

**Table 1.** Experimental Estimates of $\alpha$, $B_0$, and $\lambda$ for Samples 1 and 2 Compared to Those Computed for (13,10), (19,10), and (19,19) CNTs from the Tight-Binding Calculations

| sample | diameter, nm | length, $\mu$m | $\alpha(T > 80$ K), meV/T | $B_0$, T | $\lambda$, meV/T |
|---|---|---|---|---|---|
| 1 | 1.5 ± 0.5 | 2.8 | 2.4 ± 0.1 | 6 ± 0.5 | 0.6 ± 0.2 |
| 2 | 2.6 ± 0.5 | 5.0 | 2.0 ± 0.1 | 0 | 1.0 ± 0.2 |
| (13,10) | 1.56 | | 1.8 | 5.7 | 0.642 |
| (19,10) | 1.99 | | 2.3 | 6.24 | 0.824 |
| (19,19) | 2.58 | | 2.0 | 0 | 1.04 |

Despite its simplicity, this computational scheme reproduces the device behavior remarkably well in a wide range of temperatures and magnetic fields. The calculated $G(B)$ curves exhibit exponential behavior: $G \propto \exp(-\alpha B/k_B T)$ at $B > B_0$ At low temperatures, $k_B T \ll \varphi_1$, the temperature dependence $\alpha(T)/\lambda$ is well described by a $T^{1/2}$ power-law, which was found empirically above, with a prefactor depending only on $y_0$. The parameters $\varphi_1$ and $y_1$ can be evaluated from the $\alpha(T)/\lambda$ dependence at higher temperature.

As shown in Figure 3a,b, and Figure 4b, a very good agreement between the experiment and the simulations is obtained with values of $y_0 = 800$ nm (600 nm), $y_1 = 30$ nm, and $\varphi_1 = 60$ meV (40 meV) for sample 1 (sample 2). The fact that the values of $\varphi_1$ and $y_1$ are close for the two devices is consistent with their almost identical growth and sample preparation conditions. We also found that simulated $G(B)$ curves are almost insensitive to the value of $\varphi_0$ provided that $\varphi_0 > 200$ meV. Finally, the absolute value of $\alpha(T > 80K)$ suggests that our sample 1 is more likely a (19,10) CNT (see Table 1).

To further probe the curvature effects on the $\varepsilon_g(B)$ dependence, and in particular the band gap closing at $B \to B_0$, the effective activation energies $\Delta(B)$ were determined from the linear parts ($T > 70$ K) of the conductance Arrhenius plots for sample 1. In Figure 4c we overlay the experimental $\Delta(B)$ with the $\varepsilon_g(B)$ and $\Delta(B)$ calculated for a (19,10) CNT. The experimentally obtained $\Delta(B)$ decreases almost linearly as $B$ approaches $B_0$, and it perfectly coincides with the simulated $\Delta(B)$, confirming validity of our band structure calculations.

To summarize, we report on observation of a magnetic field induced conversion of initially metallic carbon nanotube devices into carbon nanotube field effect transistors. This effect results from the Aharonov-Bohm phenomena at the origin of a band gap opening in metallic nanotubes. Strong exponential magnetoresistance of our devices is observed up to room temperature. We show that the magnetic field controlled Schottky barriers significantly contribute to the CNT magnetoconductance, which may suggest new routes to engineer CNT-based devices characteristics. Additionally, in-depth analysis of the temperature-dependent CNT magnetotransport reveals unprecedented possibility to explore the symmetries of carbon nanotubes.

**Acknowledgments:** Authors acknowledge fruitful discussions with Yu. Pershin and assistance of A. Wade. Financial support of this work was provided by NHMFL In House Research Program, Ministerio de Educación y Ciencia under project TEC2006-13731-C02-01/MIC and NSF (DMR 0239721). The measurements were performed at the National High Magnetic Field Laboratory supported by NSF Cooperative Agreement No. DMR-0084173, by the State of Florida, and by the DOE.